# Patterns in Space: Coordinating Adhesion and Actomyosin Contractility at E-cadherin Junctions

SELWIN KAIXIANG WU AND ALPHA S. YAP

Division of Molecular Cell Biology, Institute for Molecular Bioscience, The University of Queensland, St Lucia, Queensland, Australia

## Abstract

Cadherin adhesion receptors are fundamental determinants of tissue organization in health and disease. Increasingly, we have come to appreciate that classical cadherins exert their biological actions through active cooperation with the contractile actin cytoskeleton. Rather than being passive resistors of detachment forces, cadherins can regulate the assembly and mechanics of the contractile apparatus itself. Moreover, coordinate spatial patterning of adhesion and contractility is emerging as a determinant of morphogenesis. Here we review recent developments in cadherins and actin cytoskeleton cooperativity, by focusing on E-cadherin adhesive patterning in the epithelia. Next, we discuss the underlying principles of cellular rearrangement during Drosophila germband extension and epithelial cell extrusion, as models of how planar and apical-lateral patterns of contractility organizes tissue architecture.

## INTRODUCTION

Epithelial cells are organized into sheets that form the covering layers of organs such as the colon and mammary gland. This requires cell-cell adhesion, supported notably by the E-cadherin adhesion receptor (Harrison et al., 2011). Coupling of cadherin adhesion to the actin cytoskeleton further reinforced adhesiveness (Ratheesh and Yap, 2012, Desai et al., 2013, Hong et al., 2013). Such reinforcement (Ladoux et al., 2010) is essential as epithelial contacts constantly experience mechanical stress during cellular turnover (Toyama et al., 2008, Mao et al., 2013) within an overcrowding epithelium (Farhadifar et al., 2007, Eisenhoffer et al., 2012), and growth factor- (Pollack et al., 1998) or morphogen-induced (Howard et al., 2011) remodeling.

Address correspondence to Selwin K. Wu or Alpha S. Yap, Division of Molecular Cell Biology, Institute for Molecular Bioscience, The University of Queensland, Brisbane, St. Lucia, QLD 4072, Australia. E-mail: selwin.wu@uqconnect.edu.au or a.yap@uq.edu.au

We have long appreciated that association of E-cadherin with the actin cytoskeleton plays an important role in allowing cadherin junctions to resist these mechanical stresses. Of note, E-cadherin adhesion transmits tension to shape the epithelium (Martin et al., 2010, Rauzi et al., 2008, Levayer and Lecuit, 2013, Gomez et al., 2011) and its dysregulation contributes to diseases, such as tumor progression to invasion and metastasis (Rodriguez et al., 2012, Thiery, 2002, Jeanes et al., 2008, Buda and Pignatelli, 2011, Wijnhoven et al., 2000). For many years, this was thought to reflect passive binding of cadherin/catenin complexes to cortical actin filaments, a process that led to stabilization and reinforcement of adhesion. Recent developments indicate, however, that cadherins can also control the dynamics of actomyosin cytoskeleton and can, indeed, support assembly of the junctional cytoskeleton. In this review article we consider the cellular and molecular mechanisms that allow cadherin adhesion to integrate with the contractile actin cytoskeleton. In particular, we draw attention to coordinated patterns of adhesion and contractility that serve to maintain and shape epithelia. We first outline the diversity of E-cadherin adhesion patterning in epithelial cohesion. Then, discuss the role of actomyosin network regulation in the strengthening of E-cadherin cell-cell contacts.

Finally, through comparison of polarized epithelial elongation and extrusion we delineate how adhesive and contractility patterning directs tissue assembly.

**E-CADHERIN AS AN ADHESION RECEPTOR**

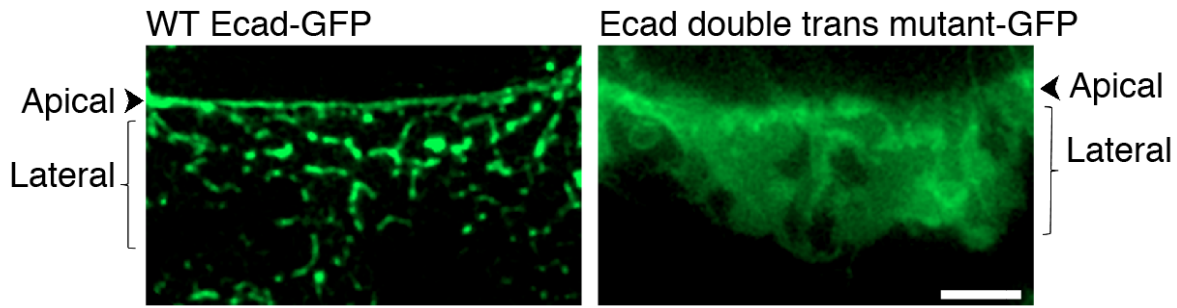

**Figure 1** Ablation of trans-interactions (W2A/K14E double mutation of EC1 domain) disrupts E-cadherin clustering at cell-cell contact. E-cadherin-GFP and Ecad-W2A/K14E mutant-GFP were expressed in Caco-2 cells depleted of endogenous E-cadherin by RNAi. Scale Bar: 5μm

**Patterning cadherin adhesion receptors within epithelial cell-cell junctions**

***Zonula adherens*** *(Figure 1)*

***Lateral Adherens Junctions*** (Figure 1 and Figure 2)

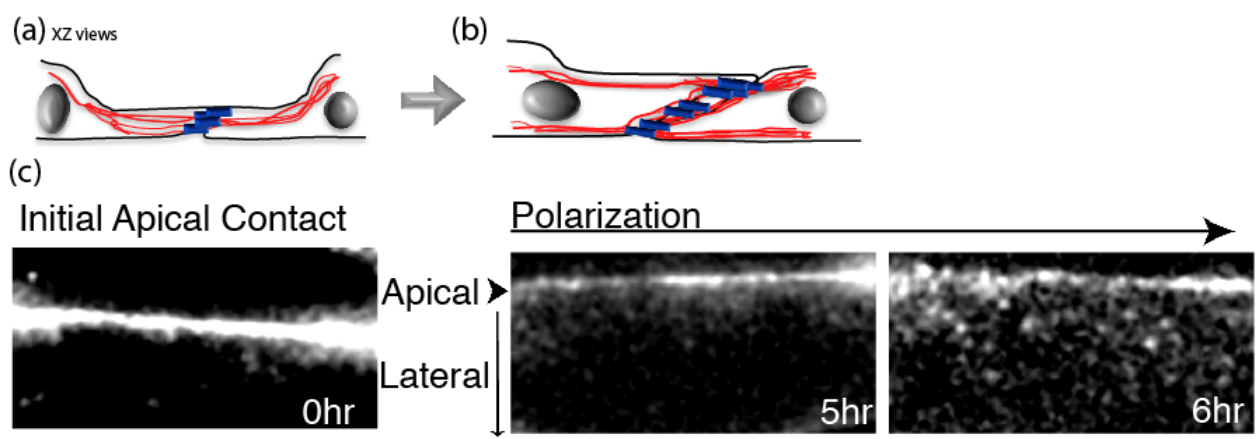

**Figure 2.** E-cad clusters distributes laterally during heightening of cell-cell contacts.

a. Initial contact formation.

b. Distribution of E-cadherin clusters laterally during lateral expansion of cell contacts forming the lateral adherens junction (LAJ).

c. LAJ formation in E-cad-GFP expressing Caco-2 cells with endogenous E-cad depleted.

***Punctate Adherens Junctions*** (Figure 3)

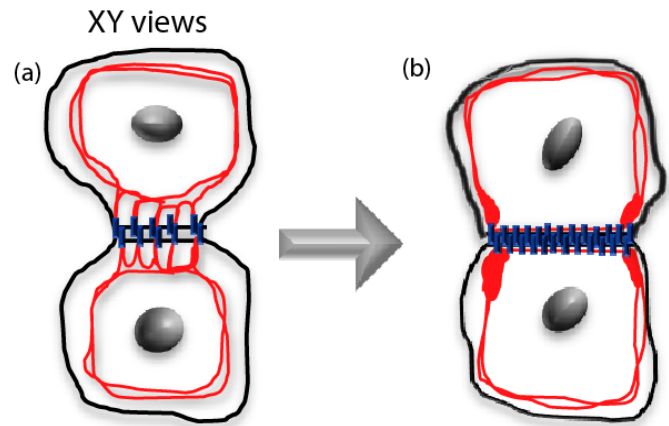

**Figure 3**. Initial contact formation process.

a. Formation of punctate adherens junctions upon initial contact of two cells.

b. Increase punctate adherens junctions (PAJ) density to establish the Zonula Adherens

***Spot adherens junctions.*** (Figure 4)

**Building a contractile apparatus at cadherin junctions.**

***Binding Cadherin complexes to cortical actin: the role of a-catenin?***

**Regulation of junctional actin dynamics.**

*i)  Cadherin junctions and actin nucleation*. actin network to junctional actin filaments
*ii) Post-nucleation regulation at cadherin junctions*.

***Building the contractile apparatus.***

**Morphogenetic implications of contractile and adhesion patterning.**

**Planar coordination of adhesion and contractility during polarized epithelia elongation: Germband Extension.** (Figure 4)

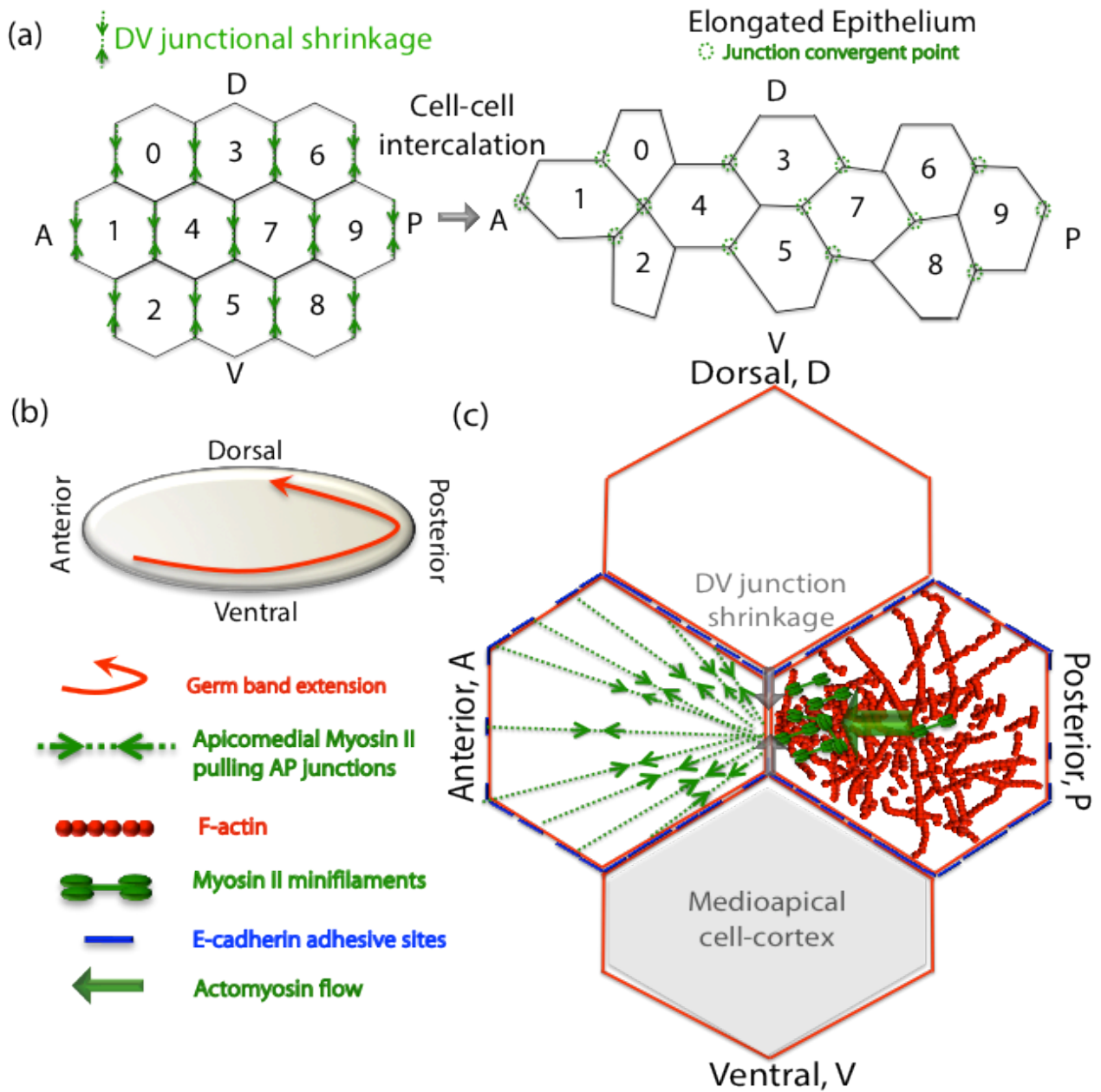

**Figure 4.** Model of Polarized Epithelial Elongation: Germband-Extension in Drosophila Embryo.

a. Shrinkage of dorsal-ventral junctions (green arrows) is a critical step in cell-cell intercalation.

b. Direction of polarized epithelia elongation in a drosophila embryo (red arrow).

c. T1 transition: Medioapical actin network anchors on DE-cadherin, where DE-cadherin distribution is planar polarized. Medioapical actomyosin contraction and its planar polarized flow into DV junction stabilize shortened DV junction length.

*Epithelial Extrusion*. (Figure 5)

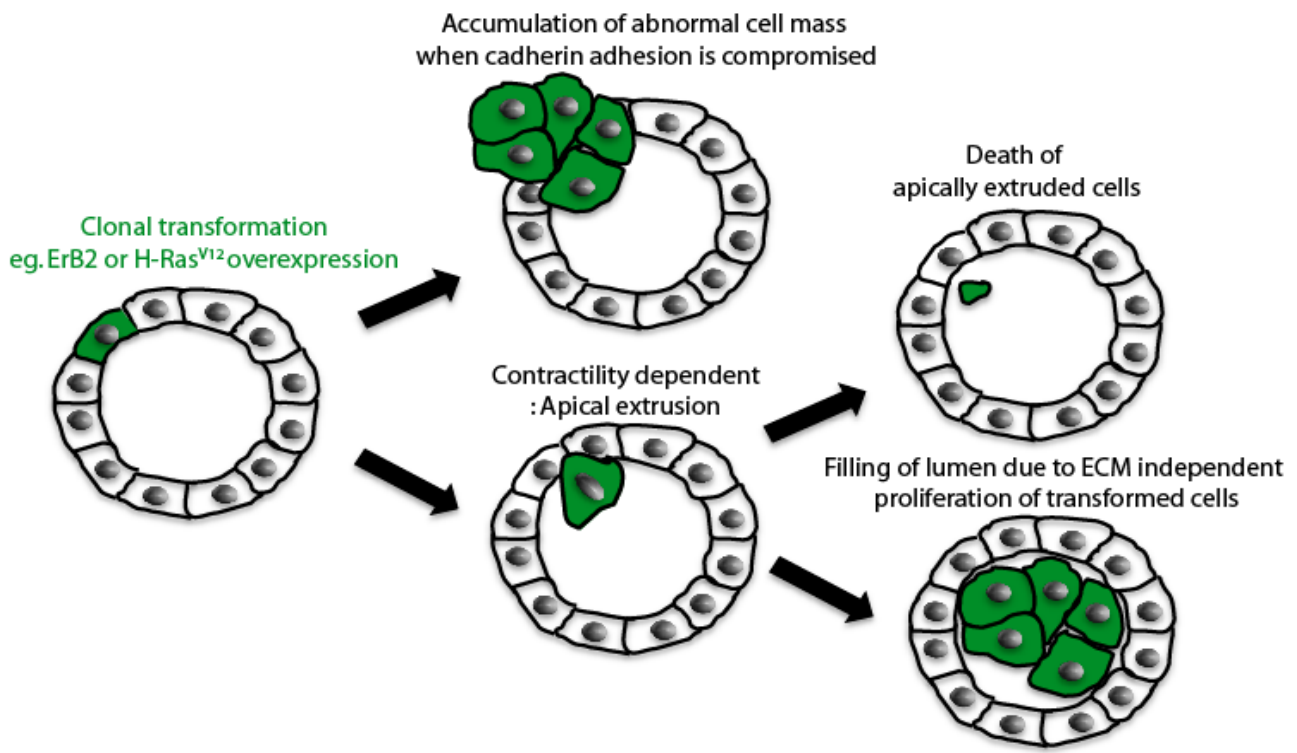

**Figure 5.** Fate of single mutated oncogenic cells (green) within a three-dimensional cyst with a central lumen and surrounded by the extracellular matrix.

**Concluding thoughts.**

The past half-decade has seen rapid progress in understanding the complex functional and mechanistic relationships between cadherin adhesion and the contractile cytoskeleton. Despite this, there is probably much more to be learnt. We would highlight two implications of the data that we have discussed. First, the potential for cadherin adhesion to regulate the actomyosin cytoskeleton implies that it may serve as an active agent in directing tissue morphogenesis. With increasing evidence that tension may affect components of the cadherin molecular apparatus, such as a-catenin (Yonemura et al., 2010), this opens the possibility that tension generated at junctions may feedback to modulate the molecular processes that are responsible for that tension (Fernandez-Gonzalez et al., 2009). Second, we propose that detailed spatial patterning in the coupling of adhesion and contractility is an important parameter that influences its morphogenetic outcome. This is highlighted by planar-polarized cooperativity between cadherin adhesion and contractility in germband extension of Drosophila. Additionally, apical-lateral patterning of contractility has been suggested to direct extrusion, yet the full extent of this phenomenon remains to be uncovered. Clearly, these are interesting issues for the future.

***Declaration of interest:*** The authors report no conflicts of interest. The authors alone are responsible for the content and writing of the paper.

**Acknowledgements:** SKW was supported by a University of Queensland Research Scholarship; ASY by grants and fellowship from the NHMRC Australia (63183, APP1010489, APP1037320). The Ecad double trans mutant was subcloned from a construct provided generously by Barry Honig (Columbia University, Howard Hughes Medical Institute).

# REFERENCES

Abe K, Takeichi M (2008). EPLIN mediates linkage of the cadherin catenin complex to F-actin and stabilizes the circumferential actin belt. *Proc Natl Acad Sci U S A*. 105: 13–19.

Aberle H, Schwartz H, Kemler R (1996). Cadherin-catenin complex: protein interactions and their implications for cadherin function. *J Cell Biochem*. 61: 514–523.

Benjamin JM, Kwiatkowski AV, Yang C, Korobova F, Pokutta S, Svitkina T, Weis WI, Nelson WJ (2010). AlphaE- catenin regulates actin dynamics independently of cadherin- mediated cell-cell adhesion. *J Cell Biol*. 189: 339–352.

Bernadskaya YY, Patel FB, Hsu HT, Soto MC (2011). Arp2/3 promotes junction formation and maintenance in the Caenorhabditis elegans intestine by regulating membrane association of apical proteins. *Mol Biol Cell*. 22: 2886–2899.

Bertet C, Sulak L, Lecuit T (2004). Myosin-dependent junction remodelling controls planar cell intercalation and axis elongation. *Nature*. 429: 667–671.

Blankenship JT, Backovic ST, Sanny JS, Weitz O, Zallen JA (2006). Multicellular rosette formation links planar cell polarity to tissue morphogenesis. *Dev Cell*. 11: 459–470.

Bryant DM, Stow JL (2004). The ins and outs of E-cadherin trafficking. *Trends Cell Biol*. 14: 427–434.

Buda A, Pignatelli M (2011). E-cadherin and the cytoskeletal network in colorectal cancer development and metastasis. *Cell Commun Adhes*. 18: 133–143.

Burstein HJ, Polyak K, Wong JS, Lester SC, Kaelin CM (2004). Ductal carcinoma in situ of the breast. *N Engl J Med*. 350: 1430–1441.

Cagan RL, Aguirre-Ghiso JA (2012). A local view of cancer. *Dev Cell*. 22: 472–474.

Carramusa L, Ballestrem C, Zilberman Y, Bershadsky AD (2007). Mammalian diaphanous-related formin Dia1 controls the organization of E-cadherin-mediated cell-cell junctions. *J Cell Sci*. 120: 3870–3882.

Cavey M, Rauzi M, Lenne PF, Lecuit T (2008). A two-tiered mechanism for stabilization and immobilization of E-cadherin. *Nature*. 453: 751–756.

Chen WC, Obrink B (1991). Cell-cell contacts mediated by E-cadherin (uvomorulin) restrict invasive behavior of L-cells. *J Cell Biol*. 114: 319–327.

Chesarone MA, Dupage AG, Goode BL (2010). Unleashing formins to remodel the actin and microtubule cytoskeletons. *Nat Rev Mol Cell Biol*. 11: 62–74.

den Elzen N, Buttery CV, Maddugoda MP, Ren G, Yap AS (2009). Cadherin adhesion receptors orient the mitotic spindle during symmetric cell division in mammalian epithelia. *Mol Biol Cell*. 20: 3740–3750.

Desai R, Sarpal R, Ishiyama N, Pellikka M, Ikura M, Tepass U (2013). Monomeric alpha-catenin links cadherin to the actin cytoskeleton. *Nat Cell Biol*. 15: 261–273.

Dominguez R, Holmes KC (2011). Actin structure and function. *Annu Rev Biophys*. 40: 169–186.

Drees F, Pokutta S, Yamada S, Nelson WJ, Weis WI (2005). Alpha-catenin is a molecular switch that binds E-cadherin- beta-catenin and regulates actin-filament assembly. *Cell*. 123: 903–915.

Eisenhoffer GT, Loftus PD, Yoshigi M, Otsuna H, Chien CB, Morcos PA, Rosenblatt J (2012). Crowding induces live cell extrusion to maintain homeostatic cell numbers in epithelia. *Nature*. 484: 546–549.

Farhadifar R, Roper JC, Aigouy B, Eaton S, Julicher F (2007). The influence of cell mechanics, cell-cell interactions, and proliferation on epithelial packing. *Curr Biol*. 17: 2095–2104.

Farquhar MG, Palade GE (1963). Junctional complexes in various epithelia. *J Cell Biol*. 17: 375–412.

Fernandez-Gonzalez R, Simoes Sde M, Roper JC, Eaton S, Zallen JA (2009). Myosin II dynamics are regulated by tension in intercalating cells. *Dev Cell*. 17: 736–743.

Fujita Y, Krause G, Scheffner M, Zechner D, Leddy HE, Behrens J, Sommer T, Birchmeier W (2002). Hakai, a c-Cbl-like protein, ubiquitinates and induces endocytosis of the E-cadherin complex. *Nat Cell Biol*. 4: 222–231.

Gomez GA, McLachlan RW, Yap AS (2011). Productive tension: force-sensing and homeostasis of cell-cell junctions. *Trends Cell Biol*. 21: 499–505.

Gournier H, Goley ED, Niederstrasser H, Trinh T, Welch MD (2001). Reconstitution of human Arp2/3 complex reveals critical roles of individual subunits in complex structure and activity. *Mol Cell*. 8: 1041–1052.

Halbleib JM, Nelson WJ (2006). Cadherins in development: cell adhesion, sorting, and tissue morphogenesis. *Genes Dev*. 20: 3199–3214.

Harrison OJ, Bahna F, Katsamba PS, Jin X, Brasch J, Vendome J, Ahlsen G, Carroll KJ, Price SR, Honig B, Shapiro L (2010). Two-step adhesive binding by classical cadherins. *Nat Struct Mol Biol*. 17: 348–357.

Harrison OJ, Jin X, Hong S, Bahna F, Ahlsen G, Brasch J, Wu Y, Vendome J, Felsovalyi K, Hampton CM, Troyanovsky RB, Ben- Shaul A, Frank J, Troyanovsky SM, Shapiro L, Honig B (2011). The extracellular architecture of adherens junctions revealed by crystal structures of type I cadherins. *Structure*. 19: 244–256.

Haviv L, Gillo D, Backouche F, Bernheim-Groswasser A (2008). A cytoskeletal demolition worker: myosin II acts as an actin depolymerization agent. *J Mol Biol*. 375: 325–330.

Herszterg S, Leibfried A, Bosveld F, Martin C, Bellaiche Y (2013). Interplay between the dividing cell and its neighbors regulates adherens junction formation during cytokinesis in epithelial tissue. *Dev Cell*. 24: 256–270.

Hinck L, Nathke IS, Papkoff J, Nelson WJ (1994). Dynamics of cadherin/catenin complex formation: novel protein interactions and pathways of complex assembly. *J Cell Biol*.125: 1327–1340.

Hirokawa N, Heuser JE (1981). Quick-freeze, deep-etch visualization of the cytoskeleton beneath surface differentiations of intestinal epithelial cells. *J Cell Biol*. 91: 399–409.Hogan C, Dupre-Crochet S, Norman M, Kajita M,

Zimmermann C, Pelling AE, Piddini E, Baena-Lopez LA, Vincent JP, Itoh Y, Hosoya H, Pichaud F, Fujita Y (2009). Characterization of the interface between normal and transformed epithelial cells. *Nat Cell Biol*. 11: 460–467.

Hong S, Troyanovsky RB, Troyanovsky SM (2010). Spontaneous assembly and active disassembly balance adherens junction homeostasis. *Proc Natl Acad Sci U S A*. 107: 3528–3533.

Hong S, Troyanovsky RB, Troyanovsky SM (2011). Cadherin exits the junction by switching its adhesive bond. *J Cell Biol*. 192: 1073–1083.

Hong S, Troyanovsky RB, Troyanovsky SM (2013). Binding to F-actin guides cadherin cluster assembly, stability, and movement. *J Cell Biol*. 201: 131–143.

Howard J, Grill SW, Bois JS (2011). Turing's next steps: the mechanochemical basis of morphogenesis. *Nat Rev Mol Cell Biol*. 12: 392–398.

Ideses Y, Sonn-Segev A, Roichman Y, Bernheim-Groswasser A (2013). Myosin II does it all: assembly, remodeling, and disassembly of actin networks are governed by myosin II activity. *Soft Matter*. 9: 7127–7137.

Imamura Y, Itoh M, Maeno Y, Tsukita S, Nagafuchi A (1999). Functional domains of alpha-catenin required for the strong state of cadherin-based cell adhesion. *J Cell Biol*. 144: 1311–1322.

Irvine KD, Wieschaus E (1994). Cell intercalation during Drosophila germband extension and its regulation by pair- rule segmentation genes. *Development*. 120: 827–841.

Jeanes A, Gottardi CJ, Yap AS (2008). Cadherins and cancer: how does cadherin dysfunction promote tumor progression? *Oncogene*. 27: 6920–6929.

Kajita M, Hogan C, Harris AR, Dupre-Crochet S, Itasaki N, Kawakami K, Charras G, Tada M, Fujita Y (2010). Interaction with surrounding normal epithelial cells influences signalling pathways and behaviour of Src-transformed cells. *J Cell Sci*. 123: 171–180.

Kametani Y, Takeichi M (2007). Basal-to-apical cadherin flow at cell junctions. *Nat Cell Biol*. 9: 92–98.

Kim SK, Shindo A, Park TJ, Oh EC, Ghosh S, Gray RS, Lewis RA, Johnson CA, Attie-Bittach T, Katsanis N, Wallingford JB (2010). Planar cell polarity acts through septins to control collective cell movement and ciliogenesis. *Science*. 329: 1337–1340.

Kobielak A, Pasolli HA, Fuchs E (2004). Mammalian formin-1 participates in adherens junctions and polymerization of linear actin cables. *Nat Cell Biol*. 6: 21–30.

Kofron M, Spagnuolo A, Klymkowsky M, Wylie C, Heasman J (1997). The roles of maternal alpha-catenin and plakoglobin in the early Xenopus embryo. *Development*. 124: 1553–1560.

Koslov ER, Maupin P, Pradhan D, Morrow JS, Rimm DL (1997). Alpha-catenin can form asymmetric homodimeric complexes and/or heterodimeric complexes with beta-catenin. *J Biol Chem*. 272: 27301–27306.

Kovacs EM, Goodwin M, Ali RG, Paterson AD, Yap AS (2002). Cadherin-directed actin assembly: E-cadherin physically associates with the Arp2/3 complex to direct actin assembly in nascent adhesive contacts. *Curr Biol*. 12: 379–382.

Kovacs EM, Verma S, Ali RG, Ratheesh A, Hamilton NA, Akhmanova A, Yap AS (2011). N-WASP regulates the epithelial junctional actin cytoskeleton through a non-canonical post-nucleation pathway. *Nat Cell Biol*. 13: 934–943.

Kuhn JR, Pollard TD (2005). Real-time measurements of actin filament polymerization by total internal reflection fluorescence microscopy. *Biophys J*. 88: 1387–1402.

Ladoux B, Anon E, Lambert M, Rabodzey A, Hersen P, Buguin A, Silberzan P, Mege RM (2010). Strength dependence of cadherin-mediated adhesions. *Biophys J*. 98: 534–542. Lambert M, Choquet D, Mege RM (2002).

Dynamics of ligand- induced, Rac1-dependent anchoring of cadherins to the actin cytoskeleton. *J Cell Biol*. 157: 469–479.

Larsson LI (2006). Distribution of E-cadherin and beta-catenin in relation to cell maturation and cell extrusion in rat andmouse small intestines. *Histochem Cell Biol*. 126: 575–582.

Leckband D, Prakasam A (2006). Mechanism and dynamics of cadherin adhesion. *Annu Rev Biomed Eng*. 8: 259–287.

Leung CT, Brugge JS (2012). Outgrowth of single oncogene- expressing cells from suppressive epithelial environments. *Nature*. 482: 410–413.

Levayer R, Lecuit T (2012). Biomechanical regulation of contractility: spatial control and dynamics. *Trends Cell Biol*. 22: 61–81.

Levayer R, Lecuit T (2013). Oscillation and Polarity of E-Cadherin

Asymmetries Control Actomyosin Flow Patterns during Morphogenesis. *Dev Cell*. 26: 162–175. Levayer R, Pelissier-Monier A, Lecuit T (2011).

Spatial regulation of Dia and Myosin-II by RhoGEF2 controls initiation of E-cadherin endocytosis during epithelial morphogenesis. *Nat Cell Biol*. 13: 529–540.

Liu R, Linardopoulou EV, Osborn GE, Parkhurst SM (2010). Formins in development: orchestrating body plan origami. *Biochim Biophys Acta*. 1803: 207–225.

Madduogoda MP, Crampton MS, Shewan AM, Yap AS (2007). Myosin VI and vinculin cooperate during the morphogenesis of cadherin cell cell contacts in mammalian epithelial cells. *J Cell Biol*. 178: 529–540.

Mangold S, Norwood SJ, Yap A, Collins BM (2012). The juxtamembrane domain of the E-cadherin cytoplasmic tail contributes to its interaction with Myosin VI. *Bioarchitecture*. 2.

Mangold S, Wu SK, Norwood SJ, Collins BM, Hamilton NA, Thorn P, Yap AS (2011). Hepatocyte growth factor acutely perturbs actin filament anchorage at the epithelial zonula adherens. *Curr Biol*. 21: 503–507.

McLachlan RW, Kraemer A, Helwani FM, Kovacs EM, Yap AS (2007). E-cadherin adhesion activates c-Src signaling at cell- cell contacts. *Mol Biol Cell*. 18: 3214–3223.

Meng W, Mushika Y, Ichii T, Takeichi M (2008). Anchorage of microtubule minus ends to adherens junctions regulates epithelial cell-cell contacts. *Cell*. 135: 948–959.

Meng W, Takeichi M (2009). Adherens junction: molecular architecture and regulation. *Cold Spring Harb Perspect Biol*. 1: a002899.

Miyaguchi K (2000). Ultrastructure of the zonula adherens revealed by rapid-freeze deep-etching. *J Struct Biol*. 132: 169–178.

Murrell MP, Gardel ML (2012). F-actin buckling coordinates contractility and severing in a biomimetic actomyosin cortex. *Proc Natl Acad Sci U S A*. 109: 20820–20825.

Nagafuchi A, Ishihara S, Tsukita S (1994). The roles of catenins in the cadherin-mediated cell adhesion: functional analysis of E-cadherin-alpha catenin fusion molecules. *J Cell Biol*. 127: 235–245.

Nishimura T, Honda H, Takeichi M (2012). Planar cell polarity links axes of spatial dynamics in neural-tube closure. *Cell*. 149: 1084–1097.

Ozawa M, Kemler R (1992). Molecular organization of the uvomorulin-catenin complex. *J Cell Biol*. 116: 989–996.

Pacquelet A, Rorth P (2005). Regulatory mechanisms required for DE-cadherin function in cell migration and other types of adhesion. *J Cell Biol*. 170: 803–812.

Perret E, Benoliel AM, Nassoy P, Pierres A, Delmas V, Thiery JP, Bongrand P, Feracci H (2002). Fast dissociation kinetics between individual E-cadherin fragments revealed by flow chamber analysis. *EMBO J*. 21: 2537–2546.

Pokutta S, Weis WI (2000). Structure of the dimerization and beta-catenin-binding region of alpha-catenin. *Mol Cell*. 5: 533–543.

Pollack AL, Runyan RB, Mostov KE (1998). Morphogenetic mechanisms of epithelial tubulogenesis: MDCK cell polarity is transiently rearranged without loss of cell-cell contact during scatter factor/hepatocyte growth factor-induced tubulogenesis. *Dev Biol*. 204: 64–79.

Pollard TD (2007). Regulation of actin filament assembly by Arp2/3 complex and formins. *Annu Rev Biophys Biomol Struct*. 36: 451–477.

Pollard TD, Blanchoin L, Mullins RD (2000). Molecular mechanisms controlling actin filament dynamics in nonmuscle cells. *Annu Rev Biophys Biomol Struct*. 29: 545–576.

Ratheesh A, Gomez GA, Priya R, Verma S, Kovacs EM, Jiang K, Brown NH, Akhmanova A, Stehbens SJ, Yap AS (2012). Centralspindlin and alpha-catenin regulate Rho signalling at the epithelial zonula adherens. *Nat Cell Biol*. 14: 818–828.

Ratheesh A, Yap AS (2012). A bigger picture: classical cadherins and the dynamic actin cytoskeleton. *Nat Rev Mol Cell Biol*. 13: 673–679.

Rauzi M, Lenne PF, Lecuit T (2010). Planar polarized actomyosin contractile flows control epithelial junction remodelling. *Nature*. 468: 1110–1114.

Rauzi M, Verant P, Lecuit T, Lenne PF (2008). Nature and anisotropy of cortical forces orienting Drosophila tissue morphogenesis. *Nat Cell Biol*. 10: 1401–1410.

Ren G, Helwani FM, Verma S, McLachlan RW, Weed SA, Yap AS (2009). Cortactin is a functional target of E-cadherin- activated Src family kinases in MCF7 epithelial monolayers. *J Biol Chem*. 284: 18913–18922.

Reymann AC, Boujemaa-Paterski R, Martiel JL, Guerin C, Cao W, Chin HF, DE La Cruz EM, Thery M, Blanchoin L

Mao Y, Tournier AL, Hoppe A, Kester L, Thompson BJ, Tapon N (2013). Differential proliferation rates generate patterns of mechanical tension that orient tissue growth. *EMBO J*. (Epub ahead of print).

Marinari E, Mehonic A, Curran S, Gale J, Duke T, Baum B (2012). Live-cell delamination counterbalances epithelial growth to limit tissue overcrowding. *Nature*. 484: 542–545.

Marshall TW, Lloyd IE, Delalande JM, Nathke I, Rosenblatt J (2011). The tumor suppressor adenomatous polyposis coli controls the direction in which a cell extrudes from an epithelium. *Mol Biol Cell*. 22: 3962–3970.

Martin AC, Gelbart M, Fernandez-Gonzalez R, Kaschube M, Wieschaus EF (2010). Integration of contractile forces during tissue invagination. *J Cell Biol*. 188: 735–749.

Martin AC, Kaschube M, Wieschaus EF (2009). Pulsed contractions of an actin-myosin network drive apical constriction. *Nature*. 457: 495–499.

Mason FM, Tworoger M, Martin AC (2013). Apical domain polarization localizes actin-myosin activity to drive ratchet- like apical constriction. *Nat Cell Biol*. 15: 926–936.

Actin network architecture can determine myosin motor activity. *Science*. 336: 1310–1314.Rimm DL, Koslov ER, Kebriaei P, Cianci CD, Morrow JS (1995).

Alpha 1(E)-catenin is an actin-binding and -bundling protein mediating the attachment of F-actin to the membrane adhesion complex. *Proc Natl Acad Sci U S A*. 92: 8813–8817.

Rizvi SA, Neidt EM, Cui J, Feiger Z, Skau CT, Gardel ML, Kozmin SA, Kovar DR (2009). Identification and characterization of a small molecule inhibitor of formin-mediated actin assembly. *Chem Biol*. 16: 1158–1168.

Rodriguez FJ, Lewis-Tuffin LJ, Anastasiadis PZ (2012). E-cadherin's dark side: possible role in tumor progression. *Biochim Biophys Acta*. 1826: 23–31.

Roh-Johnson M, Shemer G, Higgins CD, Mcclellan JH, Werts AD, Tulu US, Gao L, Betzig E, Kiehart DP, Goldstein B (2012). Triggering a cell shape change by exploiting preexisting actomyosin contractions. *Science*. 335: 1232–1235.

Rosenblatt J, Raff MC, Cramer LP (2001). An epithelial cell destined for apoptosis signals its neighbors to extrude it by an actin- and myosin-dependent mechanism. *Curr Biol*. 11: 1847–1857.

Rosin-Arbesfeld R, Ihrke G, Bienz M (2001). Actin-dependent membrane association of the APC tumour suppressor in polarized mammalian epithelial cells. *EMBO J*. 20: 5929–5939.

Rouiller I, Xu XP, Amann KJ, Egile C, Nickell S, Nicastro D, Li R, Pollard TD, Volkmann N, Hanein D (2008). The structural basis of actin filament branching by the Arp2/3 complex. *J Cell Biol*. 180: 887–895.

Sako Y, Nagafuchi A, Tsukita S, Takeichi M, Kusumi A (1998). Cytoplasmic regulation of the movement of E-cadherin on the free cell surface as studied by optical tweezers and single particle tracking: corralling and tethering by the membrane skeleton. *J Cell Biol*. 140: 1227–1240.

Saravanan S, Meghana C, Narasimha M (2013). Local, cell- nonautonomous feedback regulation of myosin dynamics patterns transitions in cell behavior: a role for tension and geometry? *Mol Biol Cell*. 24: 2350–2361.

Sarpal R, Pellikka M, Patel RR, Hui FY, Godt D, Tepass U (2012). Mutational analysis supports a core role for Drosophila alpha- catenin in adherens junction function. *J Cell Sci*. 125: 233–245.

Schlegelmilch K, Mohseni M, Kirak O, Pruszak J, Rodriguez JR, Zhou D, Kreger BT, Vasioukhin V, Avruch J, Brummelkamp TR, CamargoFD(2011).Yap1actsdownstreamofalpha-cateninto control epidermal proliferation. *Cell*. 144: 782–795.

Slattum G, Mcgee KM, Rosenblatt J (2009). P115 RhoGEF and microtubules decide the direction apoptotic cells extrude from an epithelium. *J Cell Biol*. 186: 693–702.

Smutny M, Wu SK, Gomez GA, Mangold S, Yap AS, Hamilton NA (2011). Multicomponent analysis of junctional movements regulated by myosin II isoforms at the epithelial zonula adherens. *PLoS One*. 6: e22458.

Taguchi K, Ishiuchi T, Takeichi M (2011). Mechanosensitive EPLIN-dependent remodeling of adherens junctions regulates epithelial reshaping. *J Cell Biol*. 194: 643–656.

Takeichi M (1991). Cadherin cell adhesion receptors as a morphogenetic regulator. *Science*. 251: 1451–1455.

Tang VW, Brieher WM (2012). alpha-Actinin-4/FSGS1 is required for Arp2/3-dependent actin assembly at the adherens junction. *J Cell Biol*. 196: 115–130.

Tepass U (1996). Crumbs, a component of the apical membrane, is required for zonula adherens formation in primary epithelia of Drosophila. *Dev Biol*. 177: 217–225.

Tepass U, Hartenstein V (1994). The development of cellular junctions in the Drosophila embryo. *Dev Biol*. 161: 563–596.

Thiery JP (2002). Epithelial-mesenchymal transitions in tumour progression. *Nat Rev Cancer*. 2: 442–454.

Torres M, Stoykova A, Huber O, Chowdhury K, Bonaldo P, Mansouri A, Butz S, Kemler R, Gruss P (1997). An alpha-E- catenin gene trap mutation defines its function in preimpl- antation development. *Proc Natl Acad Sci U S A*. 94: 901–906.

Toyama Y, Peralta XG, Wells AR, Kiehart DP, Edwards GS (2008). Apoptotic force and tissue dynamics during Drosophila embryogenesis. *Science*. 321: 1683–1686.

Twiss F, Le Duc Q, Van Der Horst S, Tabdili H, Van Der Krogt G, Wang N, Rehmann H, Huveneers S, Leckband DE, De Rooij J (2012). Vinculin-dependent Cadherin mechanosensing regulates efficient epithelial barrier formation. *Biol Open*. 1: 1128–1140.

Tyska MJ, Warshaw DM (2002). The myosin power stroke. *Cell Motil Cytoskeleton*. 51: 1–15.

Uyeda TQ, Iwadate Y, Umeki N, Nagasaki A, Yumura S (2011). Stretching actin filaments within cells enhances their affinity for the myosin II motor domain. *PLoS One*. 6: e26200.

Vaezi A, Bauer C, Vasioukhin V, Fuchs E (2002). Actin cable dynamics and Rho/Rock orchestrate a polarized cytoskeletal architecture in the early steps of assembling a stratified epithelium. *Dev Cell*. 3: 367–381.

Vasioukhin V, Bauer C, Yin M, Fuchs E (2000). Directed actin polymerization is the driving force for epithelial cell-cell adhesion. *Cell*. 100: 209–219.

Verma S, Han SP, Michael M, Gomez GA, Yang Z, Teasdale RD, Ratheesh A, Kovacs EM, Ali RG, Yap AS (2012). A WAVE2- Arp2/3 actin nucleator apparatus supports junctional tension at the epithelial zonula adherens. *Mol Biol Cell*. 23: 4601–4610.

Verma S, Shewan AM, Scott JA, Helwani FM, den Elzen N R, Miki H, Takenawa T, Yap AS (2004). Arp2/3 activity is necessary for efficient formation of E-cadherin adhesive contacts. *J Biol Chem*. 279: 34062–34070.

Vicente-Manzanares M, Ma X, Adelstein RS, Horwitz AR (2009). Non-muscle myosin II takes centre stage in cell adhesion and migration. *Nat Rev Mol Cell Biol*. 10: 778–790.

Watanabe N, Kato T, Fujita A, Ishizaki T, Narumiya S (1999). Cooperation between mDia1 and ROCK in Rho-induced actin reorganization. *Nat Cell Biol*. 1: 136–143.

Weaver AM, Heuser JE, Karginov AV, Lee WL, Parsons JT, Cooper JA (2002). Interaction of cortactin and N-WASp with Arp2/3 complex. *Curr Biol*. 12: 1270–1278.

Wijnhoven BP, Dinjens WN, Pignatelli M (2000). E-cadherin- catenin cell-cell adhesion complex and human cancer. *Br J Surg*. 87: 992–1005.

Yamada S, Nelson WJ (2007). Localized zones of Rho and Rac activities drive initiation and expansion of epithelial cell- cell adhesion. *J Cell Biol*. 178: 517–527.

Yamada S, Pokutta S, Drees F, Weis WI, Nelson WJ (2005). Deconstructing the cadherin-catenin-actin complex. *Cell*. 123: 889–901.

Yap AS, Brieher WM, Pruschy M, Gumbiner BM (1997). Lateral clustering of the adhesive ectodomain: a fundamental determinant of cadherin function. *Curr Biol*. 7: 308–315.

Yonemura S (2011). Cadherin-actin interactions at adherens junctions. *Curr Opin Cell Biol*. 23: 515–522.

Yonemura S, Wada Y, Watanabe T, Nagafuchi A, Shibata M (2010). alpha-Catenin as a tension transducer that induces adherens junction development. *Nat Cell Biol*. 12: 533–542.

Zhang J, Betson M, Erasmus J, Zeikos K, Bailly M, Cramer LP, Braga VM (2005). Actin at cell-cell junctions is composed of two dynamic and functional populations. *J Cell Sci*. 118: 5549–5562.

Zhang XF, Schaefer AW, Burnette DT, Schoonderwoert VT, Forscher P (2003). Rho-dependent contractile responses in the neuronal growth cone are independent of classical peripheral retrograde actin flow. *Neuron*. 40: 931–944.